\documentclass[10pt]{article}

\usepackage{amsmath}
\usepackage{amssymb}

\usepackage{graphicx}

\usepackage{cite}

\usepackage{color} 

\usepackage[labelfont=bf,labelsep=period,justification=raggedright]{caption}

\makeatletter
\renewcommand{\@biblabel}[1]{\quad#1.}
\makeatother

\date{}

\begin{document}

\begin{flushleft}
{\Large
\textbf{A combination of transcriptional and microRNA regulation improves the stability of the relative concentrations of target genes}
}
\\
Andrea Riba$^{1,\ast}$, 
Carla Bosia$^{2}$, 
Mariama El Baroudi$^{3}$,
Laura Ollino$^{1}$,
Michele Caselle$^{1}$
\\
\bf{1} Department of Physics and INFN, University of Torino, via Pietro Giuria 1, I-10125, Torino, Italy
\\
\bf{2} Human Genetics Foundation (HuGeF), via Nizza 52, I-10126, Torino, Italy
\\
\bf{3} National Research Council (CNR), Institute of Informatics and Telematics (IIT) and Institute of Clinical Physiology (IFC), Laboratory for Integrative System Medicine (LISM), Via Giuseppe Moruzzi 1, Pisa, I-56124, Italy
\\
$\ast$ E-mail: riba@to.infn.it
\end{flushleft}

\section*{Abstract}

\paragraph*{It is well known that, under suitable conditions, microRNAs are able to fine tune the relative concentration of their targets to any desired value. We show that this function is particularly effective when one of the targets is a Transcription Factor (TF) which regulates the other targets. This combination defines a new class of feed-forward loops (FFLs) in which the microRNA plays the role of master regulator. Using both deterministic and stochastic equations we show that these FFLs are indeed able not only to fine-tune the TF/target ratio to any desired value as a function of the miRNA concentration but also, thanks to the peculiar topology of the circuit, to ensures the stability of this ratio against stochastic fluctuations. These two effects are due to the interplay between the direct transcriptional regulation and the indirect TF/Target interaction due to competition of TF and target for miRNA binding (the so called "sponge effect"). We then perform a genome wide search of these FFLs in the human regulatory network and show that they are characterized by a very peculiar enrichment pattern. In particular they are strongly enriched in all the situations in which the TF and its target have to be precisely kept at the same concentration notwithstanding the environmental noise. As an example we discuss the FFL involving E2F1 as Transcription Factor, RB1 as target and miR-17 family as master regulator. These FFLs ensure a tight control of the E2F/RB ratio which in turns ensures the stability of the transition from the G0/G1 to the S phase in quiescent cells.
}

\section*{Author Summary}

Gene expression is controlled by a complex network of regulatory interactions which may be organized in two complementary subnetworks: the transcriptional one, mediated by Transcription Factors (TF), and the post-transcriptional one, in which a central role is played by microRNAs. In this paper we add a further step in the study of synergistic role of these layers of regulation: a stable fine tuning of the relative expression of target genes is obtained by a combination of transcriptional and post-transcriptional interactions, and such a combination ensures robustness against stochastic fluctuations. We show that optimal fine tuning is reached when the microRNA plays the role of master regulator and one of its targets is a TF which regulates the other microRNA targets. This combination defines a new class of feed-forward loops. We show that such circuitries are strongly enriched when the TF and its targets have to be precisely kept at the same concentration notwithstanding the environmental noise. We complete our analysis with a detailed description, using both deterministic and stochastic equations of the steady state concentrations of the genes involved in the motifs as a function of the miRNA concentration and of the miRNA-target interaction strength.

\section*{Introduction}

Much interest has been attracted in the last few years by the study of the 
mixed regulatory network combining transcriptional and post-transcriptional regulatory interactions (see \cite{Martinez09} for a review).
In analogy to what was done a few years ago for the purely transcriptional network \cite{Alon07} also in this more complex case
a few network motifs showing a significant topological enrichment were identified \cite{Re09, Shalgi07, Tsang07, Yu08} and their functions were studied using both
deterministic and stochastic approaches \cite{Shalgi07, Tsang07, Hornstein06, Osella11, Bosia12}

So far, the attention was mainly devoted to circuits in which miRNAs have only an auxiliary role while the role of master regulator is played by a Transcription Factor
(TF). This is the case for instance of the miRNA-mediated Feed Forward Loops (FFL) studied in \cite{Shalgi07, Tsang07, Hornstein06, Osella11} or the miRNA mediated
self-loop \cite{Bosia12}.

However it was recently realized that several important biological processes are actually controlled by miRNAs which play themselves the role of master 
regulators and that the corresponding network motifs show a remarkable degree of topological enrichment in the mixed regulatory network \cite{gerstein2012architecture, Sun12}.
From a systems biology point of view a major reason of interest of this type of  regulatory topology is the so called "sponge effect" 
\cite{arvey2010target,Salmena2011, Sumazin2011} i.e. the appearance of an indirect interaction between the
targets due to competition for miRNA binding. 

In \cite{gerstein2012architecture} analysis of data from the {\it Encyclopedia of DNA Elements (ENCODE)} 
project revealed that two distinct classes of miRNA controlled motifs  were particularly enriched in the network.
The first class contained motifs in which a miRNA targets two genes which physically interact (for example they can dimerize). 
In the second class the miRNA targets two transcription factors which both regulate the same gene, one as proximal and one as distal regulator. 
This same topology was found to be over-represented
in human glioblastoma combining bioinformatical analysis and expression data \cite{Sun12}.
Both these examples suggest a role of miRNAs in ensuring the stability and fine-tuning of the {\bf relative concentration} of their targets.

\begin{figure}[h]
	\centering
	\includegraphics{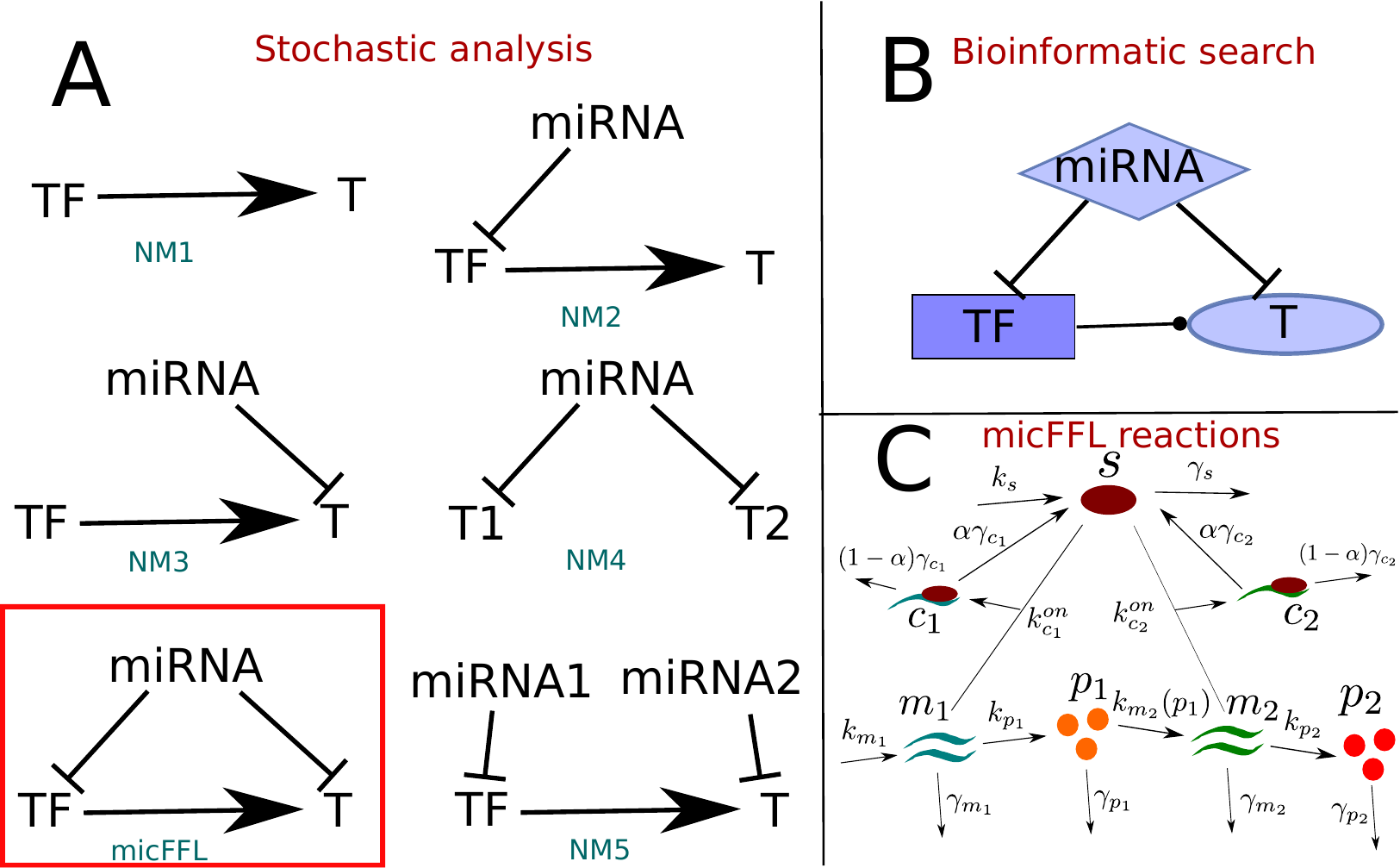}
	\caption{\label{figFFL}}
\end{figure}

Remarkably enough, it turns out that this topological enrichment is further magnified (see our analysis below) 
if one selects among these motifs those in which the two targets are linked by a
transcriptional regulation (see Figure \ref{figFFL}B). The resulting network motif is a feedforward loop in which the role of master regulator is played by a miRNA which
regulates a Transcription Factor (TF in the following) and together with it one or more target genes (T in the following). 
We shall denote in the following these motifs as "miRNA controlled FeedForward loops" (micFFL).

An interesting feature of the micFFL is that it is the simplest motif in which a 
TF regulates its target simultaneously with a direct (transcriptional)
and indirect (post-transcriptional: mediated by the sponge effect) regulatory interaction. Depending on the sign of the transcriptional regulation this 
combination can be coherent or incoherent and may have potentially very interesting functional roles.

In this paper we shall in particular address the case in which the transcriptional regulation is of activatory type and acts thus coherently with the indirect sponge
interaction (see the left bottom motif of Figure \ref{figFFL}A). As we shall see such a circuit is able to perform, depending on the values of the parameters, 
a few important functions which may enhance the cooordination of the targets action but at the same time, in some cases, may also represent a too strong linkage between them and thus decrease the overall flexibility of the network. This non-trivial behaviour could be the reason of the quite peculiar pattern of topological enrichment that we observe. 

Our main goal in this paper will be to study in a quantitative way these functions, to fix the range of parameters in which they occur
and, possibly, to understand their role within the regulatory network as a whole. We shall address the model both at the deterministic and at the stochastic level. In order to quantify the behaviour of the various molecular species involved, we shall compare the circuit with four other miRNA mediated regulatory circuits with the same players (one miRNA, one TF and one target) but different network topologies. 

In all cases the miRNA-target interaction will be modelled via a titration-like mechanism, i.e. we assumed that
miRNA and target may only interact by forming a 
complex which eventually degrades \cite{vanoudenaarden2011}. After degradation of the complex the miRNA might be recycled. 
This choice, which is at the basis of the 
 sponge effect, will play a major role in our analysis. In fact
it has been shown that titration-like mechanisms entails, among other properties, cross-talk and statistical correlation between different targets 
in competition for the same group of molecules \cite{Elf03, Levine07, Mitarai07, BosiaCerna12, Figliuzzi12}. In our particular case we will show 
that the sponge interaction between the TF and its target (i.e. their competition for miRNA-binding) induces a statistical correlation between them
much stronger than in the case of a simple transcriptional regulation, that this linkage holds for a range of miRNA concentrations larger than in the other circuits
and reaches its maximum exactly when  TF and target show the highest degree of stochastic fluctuations. 

Altogether these observations support the general picture of miRNAs as homeostasis controllers 
\cite{Sumazin2011,Osella11}, with different roles depending on the 
particular topologies they are embedded in. 
In particular, coherent micFFL could be useful in situations in which the TF and its target have to be precisely kept at the same concentration notwithstanding the 
environmental noise. 
We shall discuss in the last section a prototypical example of this situation, i.e. the 
micFFL  involving E2F1 as TF,  RB1 as target and a set of miRNAs (miR-106a, miR-106b, miR-17, miR-20a and miR-23b) as master regulators. This circuit is involved in the fine tuned control of the transition from the G0/G1 to S phase  in the cell cycle. This transition is triggered by the difference in concentration of the two targets and we shall argue below
that the micFFLs controlling the two genes were selected by evolution exactly to avoid accidental triggering of the transition due to uncorrelated stochastic fluctations of
the two proteins.
   
The comparison with the other topologies shows that the simple
loss or addition of one of the interactions in the loop could destroy this linkage and lead to pathological behaviours.

\section*{Results}

\subsection*{Bioinformatic search of micFFL in the human Regulatory Network}

A detailed description of our procedure is reported in the Material and Methods section, we only report here the main steps.
Briefly, we constructed a list of putative micFFL combining miRNA-target and TF-target regulatory interactions obtained as follows. For the miRNA-target side we integrated information obtained from four freely available databases of miRNA-target interactions, chosen so as to have the widest possible spectrum of different prediction strategies: doRiNA \cite{Anders2012}, microRNA.org \cite{Betel2008}, TargetScan \cite{Lewis2003} and PITA \cite{Kertesz2007}. We selected as potential targets only transcripts corresponding to  protein-coding genes completely annotated in Ensembl 68 \cite{Flicek2012}. For the TF-target side we used two different strategies. In the first one we selected the TFs contained in the JASPAR database \cite{Stormo2000,Wasserman2004} and used the corresponding Position Frequency Matrix (PFM) to construct a search algorithm for transcription factor binding sites (TFBS) within the target promoter regions. We found in this way a total of 948125 interactions. In the second approach we simply used as signatures of a TF-target interaction the ChIP-seq results of the ENCODE project\cite{gerstein2012architecture}. Combining together the results of the five cell lines of the ENCODE project we obtained a total of 45328 TF-target interactions. We obtained in this way a total of 75933600 micFFLs with miRNA target interaction confirmed by at least one database in the JASPAR case and a total of 2426300 micFFLs in the ENCODE case. 
We chose this twofold strategy to construct the TF-Target side of our network so as to have an independent check of our  enrichment analysis. In fact, while the ENCODE list which is based on direct experimental evidence has the obvious advantage of minimizing false positives results, it could also induce a statistical bias in the results due to the fact that ChIP-seq results are not uniformly distributed among all possible TFs and targets. This could in principle create problems when
performing a topological enrichment analysis. For this reason we chose to supplement this analysis with an alternative procedure which had exactly the opposite features: it is an unbiased genome-wide bioinformatic search from sequence information only, with no reference to experimental results. The obvious drawback of this second approach is the possible presence of several false positives in our results. As we shall see below our enrichment analysis gives similar enrichment scores for both strategies thus strongly supporting the reliability of our results. 

\subsection*{Enrichment test}
\begin{figure}[h]
	\centering
	\includegraphics{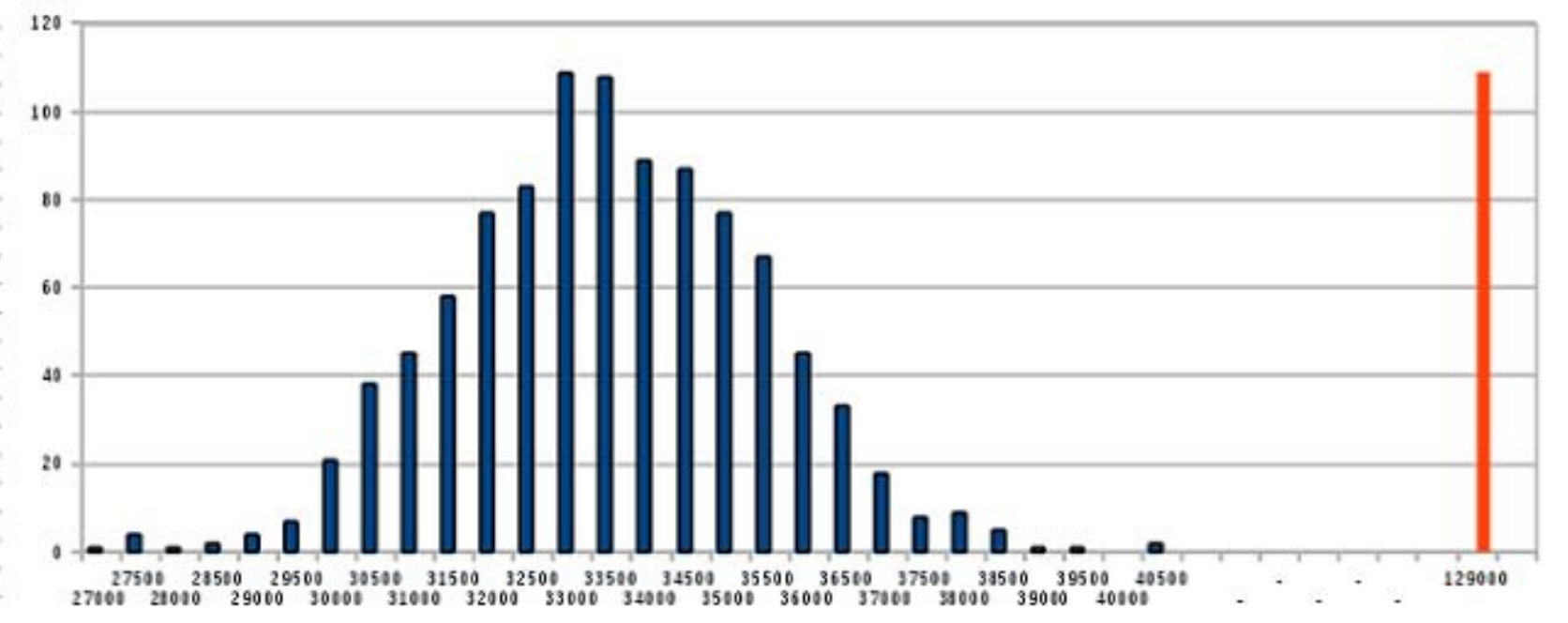}
	\caption{\label{figzA}}
\end{figure}

\begin{figure}[h]
	\centering
	\includegraphics{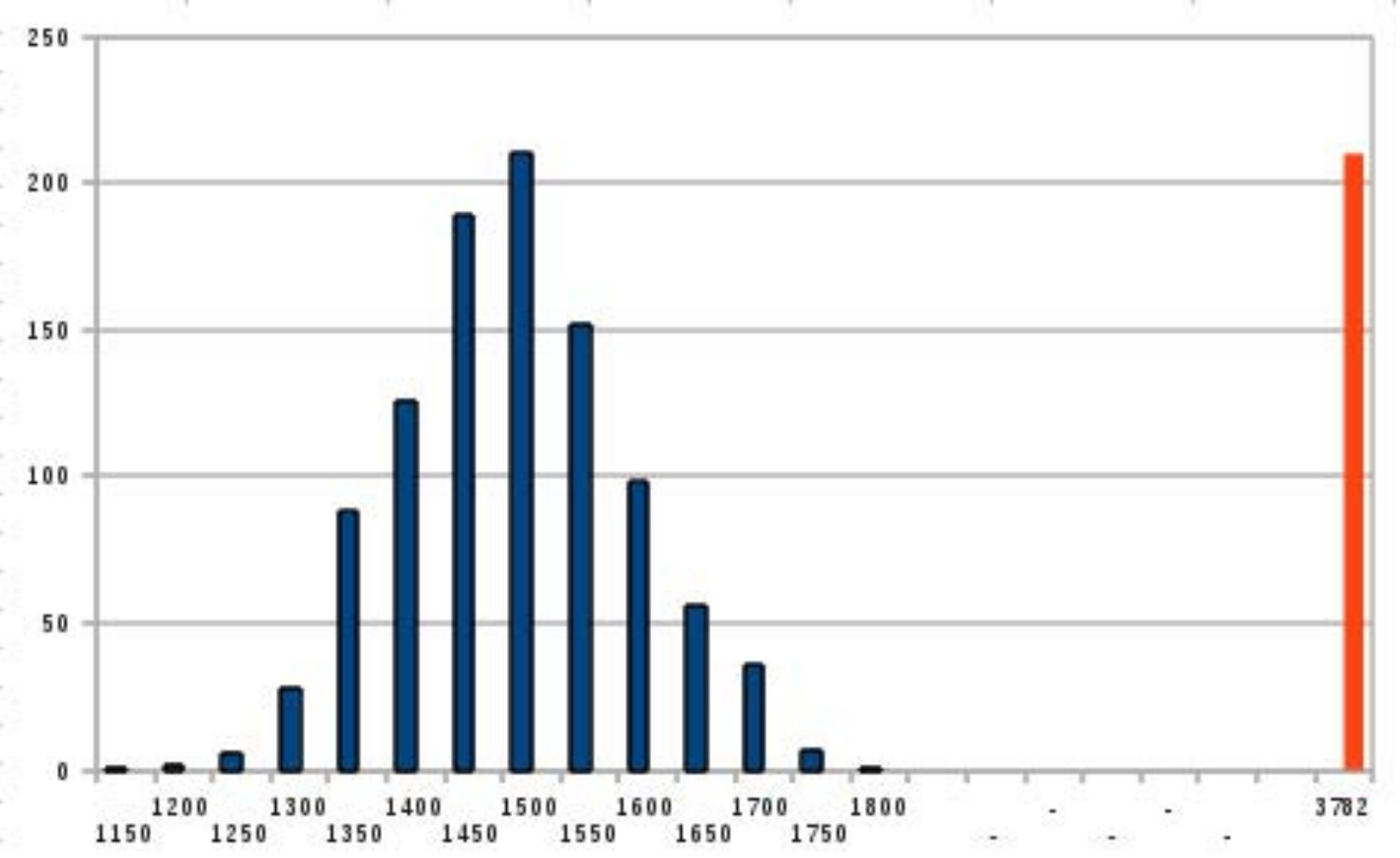}
	\caption{\label{figzB}}
\end{figure}

\begin{figure}[h]
	\centering
	\includegraphics{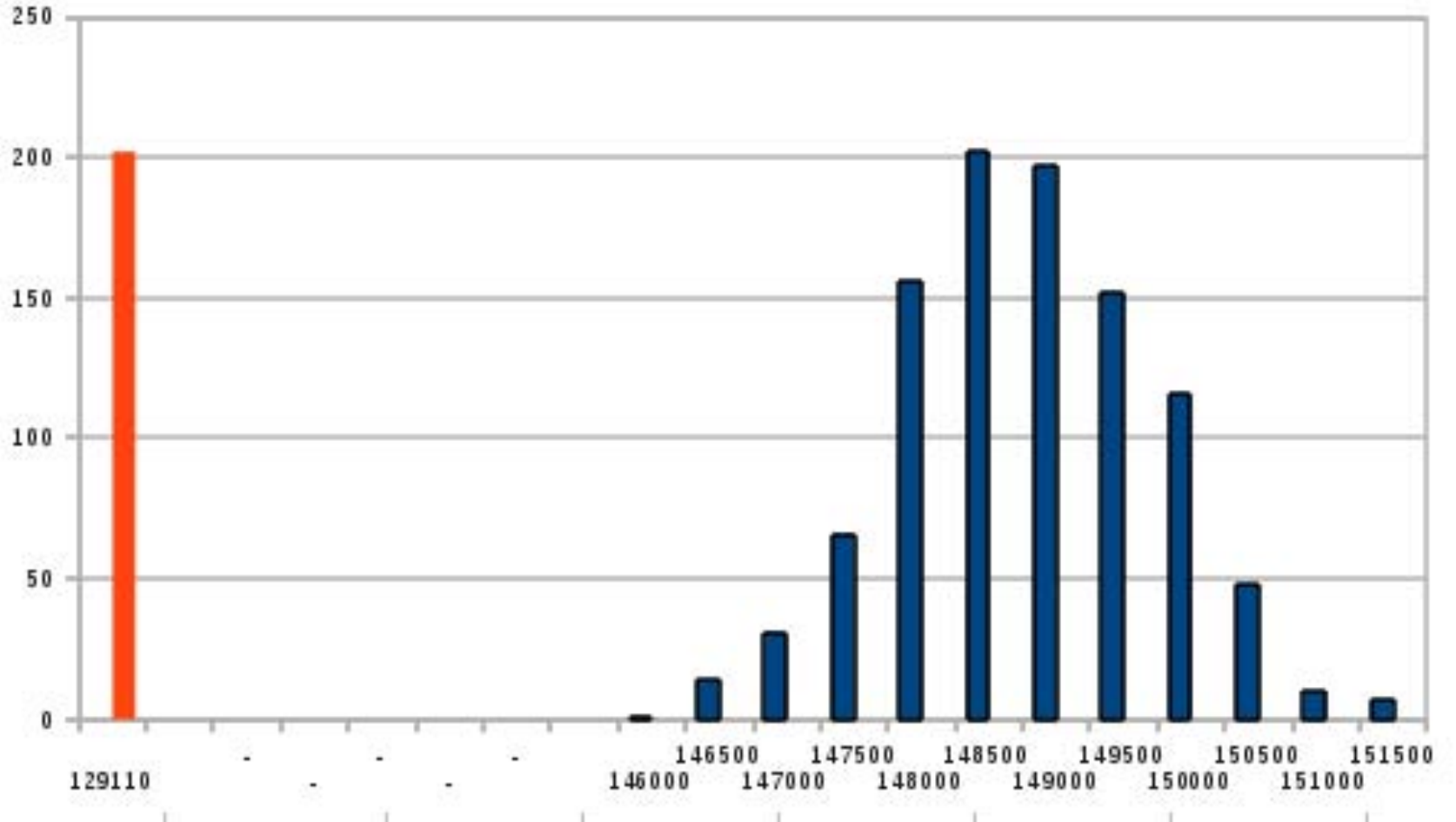}
	\caption{\label{figzC}}
\end{figure}

\begin{figure}[h]
	\centering
	\includegraphics{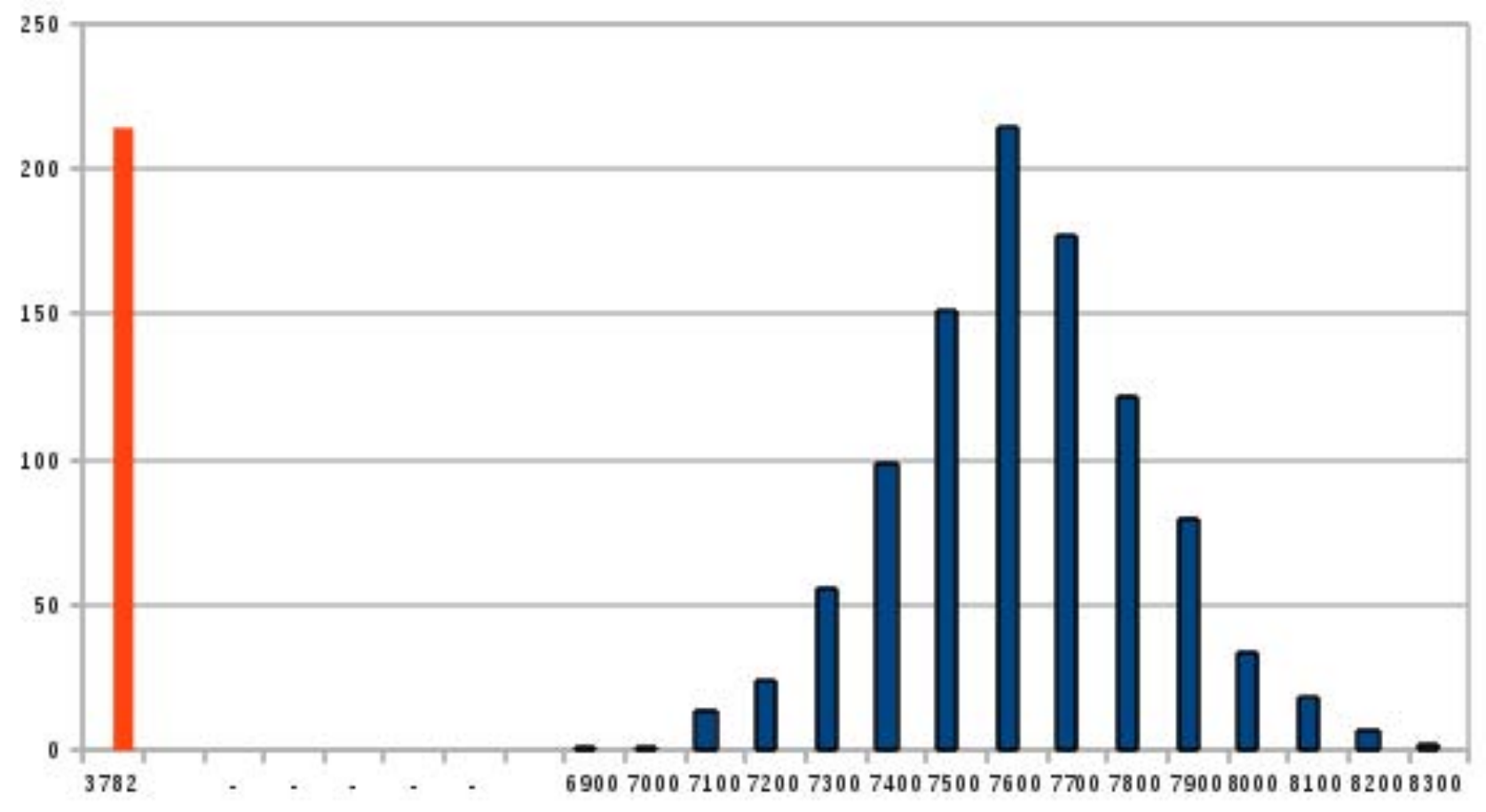}
	\caption{\label{figzD}}
\end{figure}

In order to minimize the number of false positives we selected only micFFLs in which both the miRNA-TF and the miRNA-target links were confirmed by all the four databases. This choice reduced the number of micFFLs to 129110 in the Jaspar case and 3782 in the Encode case. Since the links of the loop are not on the same ground we performed a topological enrichment analysis by random reshuffling {\sl separately} the post-transriptional and transcriptional links of the micFFL. First we randomized miRNA-target links keeping TF-target links fixed. We made 1000 simulations. For each miRNA we extracted casual targets within 
Ensembl 68 list of known protein coding transcripts keeping fixed the number of targets (i.e. keeping unchanged the outdegree of the miRNA nodes), we performed the simulation both for the JASPAR and for the ENCODE lists of TF-target interactions. In both cases we found very high values of the z-score (see Figures \ref{figzA} and \ref{figzB}): 49.4 for Jaspar and 23.3 for Encode.
Then we randomized TF-target links, keeping the miRNA-target links unchanged. Also in this case we kept fixed the outdegree of the TF nodes of the network and perfomed the reshuffling both for the JASPAR and ENCODE lists. Remarkably enough we found this time in both  cases a very strong {\sl negative} enrichment, (see Figures \ref{figzC} and \ref{figzD}) with z-score values of the same magnitude than in the previous case: -20.8 for Jaspar and -18.1 for Encode. The simplest explanation of this very peculiar behaviour is that miRNAs seem to target preferentially TFs (this largely explains the large positive enrichment in the first reshuffling test) but at the same time the particular topology of the micFFL seems to be strongly selected against by evolution and is preferentially {\sl avoided} within the network. These observations make micFFLs a very interesting subject of study. It seems that the particular topology of the micFFL induces very strong constrains on the behaviour of its targets and might be in general dangerous for the performances of the network. They also mean that when one of this circuits is actually realized in the network it is certainly not by chance and it is likely to play a well precise functional role. The remaining part of this paper will be mainly devoted to understand this issue. It is very interesting to observe that the enrichment pattern is essentially the same both in the JASPAR and in the ENCODE cases. Since the two TF datasets have a rather small overlap (only 38 TFs are in common) and the approaches to detect regulatory interactions are completely independent, the similarity of the two enrichment patterns is a strong evidence of their reliability and robustness. Finally it is worthwhile to stress that this very peculiar enrichment pattern almost disappears and would escape detection if one simultaneously permutes both transcriptional and post-transcriptional interactions due to the compensation between positive and negative enrichemnts. 

\subsection*{Putative functions of micFFLs}
It has been recently shown that microRNAs can generate thresholds in target gene expression \cite{vanoudenaarden2011} and that these thresholds can generate a sensitive response to mRNA concentration, i.e. that they induce a non-linear relation between protein and transcript concentrations. In the same paper it was also pointed out that gene expression shows large cell to cell fluctuations in a population of identically prepared cells. We shall show that similar threshold effects are also present both in the TF and in the target of micFFLs and that as a function of miRNA concentration the relative concentrations of TF and target can be fine tuned to any desired value. What is more important we shall also show that the peculiar topology of the micFFL ensures a tight control of stochastic fluctuations of this ratio and that this noise reduction is maximal exactly in the vicinity of the threshold region. We shall perform our analysis in three steps of increasing level of complexity. First we shall drastically simplify the problem, assuming a logical approximation for the Hill function involved in the transcriptional regulation, assuming equal values for the parameters controlling miRNA-TF and miRNA-target interactions and taking the limit of infinite interaction strength between the miRNA and its targets. This simplified version of the problem will allow us to have a concrete intuition of the behaviour of the circuit in the whole range of parameters. Then in the second step we shall relax these constraints. The model can be still solved exactly, the complexity of the solution increases but instead of looking at all the possible features of the model we shall only concentrate on the behaviour of the ratio between the concentration $p1/p2$ of the two targets. As we mentioned above the robustness of this ratio against stochastic fluctuations is one of the main reasons of interest of this circuit and will be the main issue of the third level of analysis in which we shall address the problem using stochastic equations.

In all the three steps, in order to discuss the functional properties of the micFFL we shall compare it with the five "null models" obtained eliminating miRNA-TF and/or miRNA-target interactions. We shall thus be able to identify which properties are direct consequences of the miRNA interaction (as for instance the threshold effect) or are a peculiar consequence of the micFFL topology (as the sponge interaction mentioned above or the noise reduction effect which we shall discuss below). 

\begin{itemize}

\item
The simplest null model is represented by the simple direct regulation TF $\rightarrow$ T without miRNAs (NM1). Comparison
 with this null model will show the effect of switching on the miRNA in our circuit.

\item
Two other important null models are those network motifs in which we only keep the miRNA-TF interaction (NM2)
 or the miRNA-target interaction (NM3) (see Figure \ref{figFFL}A). 

\item
Finally, in the next section on the stochastic analysis we shall also use for comparison two other null models: The one in which a miRNA regulates separately the two targets T1 and T2, but no regulatory interaction exists between the targets (NM4), and the open circuit in which two independent miRNAs regulate TF and target respectively (NM5), (see Figure \ref{figFFL}A).

 \end{itemize}
     
As we shall see these circuits are themselves very interesting. In particular NM4 was widely studied in the past few years to model bacterial small RNA (sRNA)/target interaction \cite{Levine07, Mitarai07} and more recently it was also discussed in the framework of a miRNA/target interaction network (\cite{BosiaCerna12, Figliuzzi12}) as an example of the so called "sponge effect".

As a byproduct of our analysis we shall also be able to discuss a few interesting features of these null models.

\subsection*{Deterministic analysis}

The micFFL is described by the following set of equations:
\begin{equation}
\begin{array}{ll}
\frac{d\,m_1}{dt} &= k_{m_1}-\gamma_{m_1}m_1-k^{on}_1 m_1 M_{free}+k^{off}_1 c_1\\
\frac{d\,p_1}{dt} &= k_{p_1}m_1-\gamma_{p_1}p_1\\
\frac{d\,m_2}{dt} &= k_{m_2}f(p_1)-\gamma_{m_2}m_2-k^{on}_2 m_2 M_{free}+k^{off}_2 c_2\\
\frac{d\,p_2}{dt} &= k_{p_2}m_2-\gamma_{p_2}p_2\\
\frac{d\,M_{free}}{dt} &= k_s - \gamma_sM_{free} - k^{on}_1m_1 M_{free} + (k^{off}_1+\gamma_{c_1})c_1 - k^{on}_2m_2 M_{free} + (k^{off}_2+\gamma_{c_2})c_2\\
\frac{d\,c_1}{dt} &= k^{on}_1m_1 M_{free} - (k^{off}_1+\gamma_{c_1})c_1\\
\frac{d\,c_2}{dt} &= k^{on}_2m_2 M_{free} - (k^{off}_2+\gamma_{c_2})c_2\\
\end{array}
\label{eq0}
\end{equation}

where $\gamma_x$ denotes the degradation constant of the molecular species $x$ and  $k_x$ the corresponding production rate; $m_1$ and $p_1$ the concentration of the TF mRNA and protein and  $m_2,p_2$ those of the target. Following \cite{vanoudenaarden2011} we shall assume a titration-like  miRNA-target interaction, i.e. we shall assume that the miRNA can interact with the target mRNA only by forming a complex $c_i$ with the mRNA $m_i$ whose stability is determined by the costants $k^{on}_i$ and $k^{off}_i$ and by the concentration of unbound miRNA which we denote as $M_{free}$. $M_{free}$ is related to the total concentration of miRNA $M_{tot}$ by:
\begin{equation}
M_{tot} = M_{free}+c_1+c_2~~~.
\end{equation}
We shall assume in the following $M_{tot}$ as an external input of our circuit. 
In the above equation we describe, as usual, the transcriptional regulation of $m_2$ using the activatory Hill function
\begin{equation}
f(p_1)=\frac{p^n_1}{p_1^n+h^n}
\end{equation}
with Hill coefficient $n$ and activation coefficient $h$.
The equations describing the null models introduced above can be easily obtained form eq.s \ref{eq0}, eliminating some of the molecular species and/or 
of the interactions. They are discussed in detail in the Supplementary Material (SM).

\subsubsection*{Steady state analysis with the logic approximation}

As a first step let as approximate the Hill function with  the Heaviside step function $f(p_1)=H(p_1-h)$. We are interested in the steady state solution of the
circuit. The equations for $p_1$ and $p_2$ can be solved immediately leading to the steady state values: $p_1^0=k_{p_1}m_1/\gamma_{p_1}$ 
and $p_2^0=k_{p_2}m_2/\gamma_{p_2}$.
We can rescale the activation coefficient $h_s \equiv h\gamma_{p_1}/k_{p_1} $
so as to write the step function as a function of $m_1$ and eliminate $p_1$ from the equations.
Following (\cite{vanoudenaarden2011}) let us introduce the quantities ($i=1,2$):

\begin{equation}
\begin{array}{l}
\lambda_i \equiv \frac{k_i^{off}+\gamma_{c_i}}{k_i^{on}}\\
\theta_i \equiv  \frac{\gamma_{c_i}}{\gamma_{m_i}}M_{tot}\\
\end{array}
\end{equation}

which have an immediate physical interpretation: $\theta_i$ is the (suitably rescaled) amount of miRNA acting on $m_i$ and $1/\lambda_i$ measures the "strenght" of this
interaction i.e. the lifetime of the complex
$c_i$. These will be in the following the only external parameters of our circuit. 
One of the goals of our analysis will be to discuss the behaviour of the various
concentration as a function of $\theta_i,\lambda_i$.

Finally let us assume for simplicity 
$\lambda_1=\lambda_2=\lambda$, $\theta_1=\theta_2=\theta$, 
and let us denote as $m_i^0$ the steady state value which $m_i$ would reach if $M_{tot}=0$ i.e. 
$m_1^0\equiv k_{m_1}/\gamma_{m_1}$ and $m_2^0\equiv k_{m_2}/\gamma_{m_2}$ if $m_1^0>h_s$ and $m_2^0=0$ otherwise.

Then it is easy to obtain the steady state values of $m_1$ e $m_2$ as a function of $\theta$ and $\lambda$.

\begin{equation}
\begin{array}{l}
{m_1}=m_1^0\frac{{m^0_1}+ m^0_2-\theta-\lambda +\sqrt{\left((m^0_1+m^0_2-\theta -\lambda )^2+4 (m^0_1+m^0_2) \lambda \right)}}{2 (m^0_1+m^0_2)}\\
{m_2}=m_2^0\frac{{m^0_1}+ m^0_2-\theta-\lambda +\sqrt{\left((m^0_1+m^0_2-\theta -\lambda )^2+4 (m^0_1+m^0_2) \lambda \right)}}{2 (m^0_1+m^0_2)}\\
\end{array}
\end{equation}

The implications of this result can be better appreciated if we take the $\lambda\to0$ limit. 

\begin{equation}
\begin{array}{l}
m_1=\left\{
	\begin{array}{ll}
	0 & m^0_1+m^0_2\leq\theta \\
	m^0_1\left(1-\frac{\theta}{m^0_1+m^0_2}\right) & m^0_1+m^0_2>\theta \\
	\end{array}\right.\\
m_2=\left\{
	\begin{array}{ll}
	0 & m^0_1+m^0_2\leq\theta \bigvee m_1<h_s \\
	m^0_2\left(1-\frac{\theta}{m^0_1+m^0_2}\right) & m^0_1+m^0_2>\theta \bigwedge m_1>h_s \\
	\end{array}\right.\\
\text{with }m^0_2=\left\{
	\begin{array}{ll}
	0 & m_1<h_s \\
	m^0_2 & m_1>h_s \\
	\end{array}\right.\\
\end{array}
\end{equation}

\begin{figure}[h]
	\centering
	\includegraphics{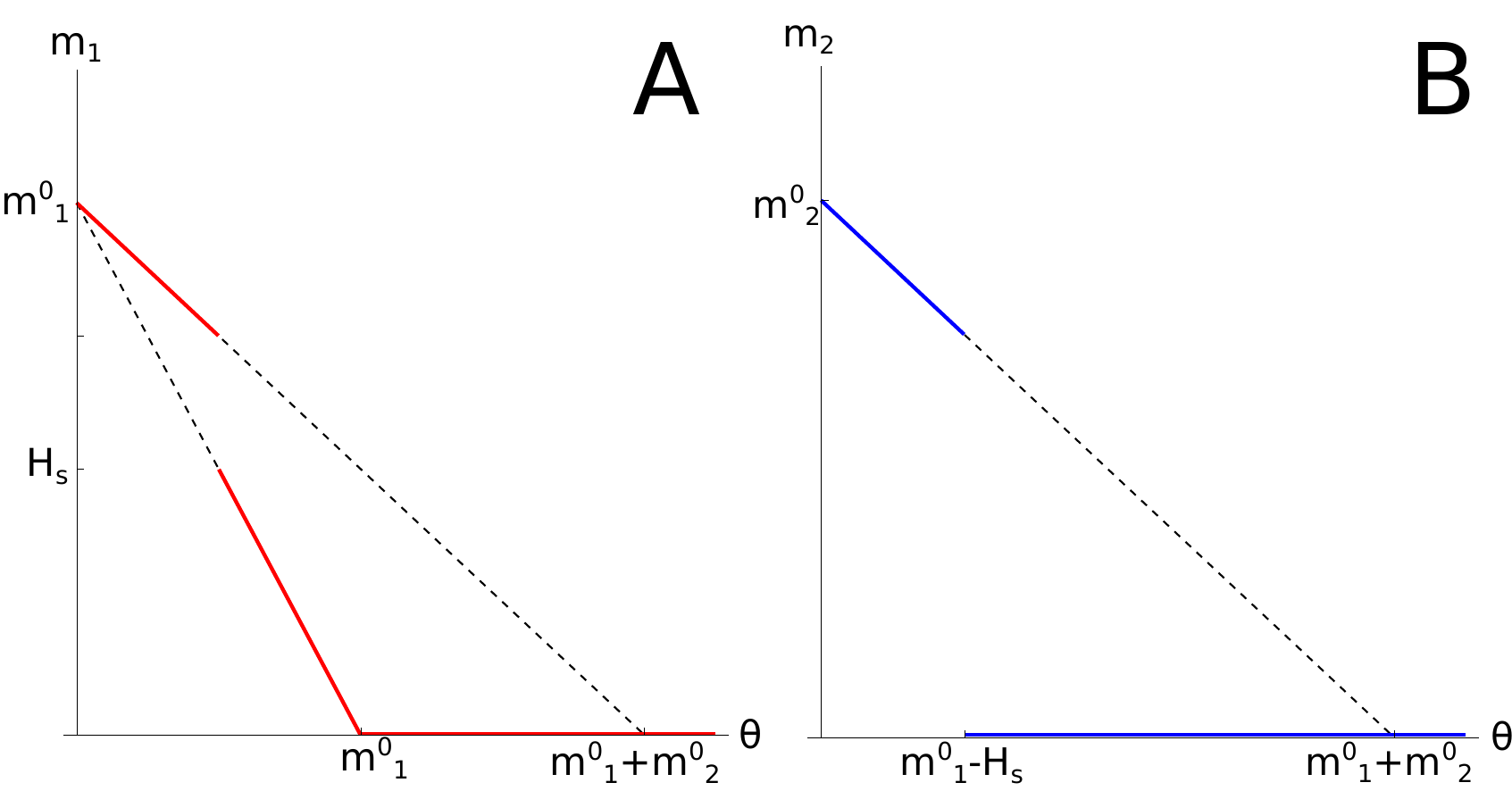}
	\caption{\label{fig:plots}} 
\end{figure}

We plot the value of $m_1$ and $m_2$ as a function of $\theta$ (i.e. of the miRNA concentration)  in Figures \ref{fig:plots}A and \ref{fig:plots}B.

Looking at these figures we see a few interesting and non trivial features:
\begin{itemize}
\item In the $\lambda\to0$ limit we find for the transcription factor $m_1$ the same  threshold behaviour discussed in \cite{vanoudenaarden2011} as a function of the
miRNA concentration. The same effect should be present also in the target concentration $m_2$, but is hidden by the fictious step behaviour due to 
the logic approximation.
It is easy to understand the origin of this threshold behaviour: 
if the number of free miRNA molecules greatly exceeds the number of transcripts mTF and mT, then these will be almost all bound in complexes and 
the corresponding proteins will not be expressed. On the opposite side, if the number of mTF and mT molecules overcomes miRNA amount, then nearly all miRNAs 
will be bound in complexes but there will be a sufficient amount of free mTFs and mTs to be translated.
\item As the total miRNA concentration decreases
the TF concentration increases following the trajectory plotted in Figure \ref{fig:plots}A. When the TF concentration reaches the
threshold $h_s$ for the  $m_2$ activation we observe a sudden enhancement in the TF concentration due to the sponge interaction  between $m_1$
and $m_2$. In fact when also $m_2$ is present then the two mRNAs start to compete for the same miRNAs and as a net effect there is a smaller amount of miRNA available to
downregulate $m_1$. This non linear behaviour of the TF concentration as the miRNA concentration increases 
is in our opinion one of the most effective ways to detect sponge-like interactions. 
\item The ratio $m_2/m_1$ (and thus obviously the ratio $p_2/p_1$) can only take two possible values: $m_2/m_1=m_2^0/m_1^0$ for $m_1>h_s$ and $m_2/m_1=0$ for $m_1<h_s$. However
 this is clearly an artifact of the logic  approximation. The correct behaviour of this ratio will be recovered in the next section.
\end{itemize}

\subsubsection*{Steady state analysis with the Hill function}
Eq.s (\ref{eq0}) can be solved in the steady state limit even without resorting to the logic approximation. 
The solution can be written in a rather simple way as a function
of $M_{free}$. Let us introduce:
\begin{equation}
\theta^{free}_i \equiv \frac{\gamma_{c_i}}{\gamma_{m_i}}M_{free},~~(i=1,2) ~~~~\eta\equiv \frac{h}{p_1^0}\left(1+\frac{\theta^{free}_1}{\lambda_1}\right)
\end{equation}
then we may write:
\begin{equation}
\begin{array}{ll}
p_1 =p_1^0 \frac{1}{1+{\theta^{free}_1}/\lambda_1} \\
p_2 =p_2^0 \frac{1}{1+{\theta^{free}_2}/\lambda_2}\frac{1}{1+\eta^{n}}\\
\end{array}
\end{equation}
where $p_1^0$ and $p_2^0$ denote the asymptotic values of $p_1$ and $p_2$ in absence of miRNAs and with the Hill function at saturation: $f(p_1)=1$,
i.e.  $p_i^0=k_{p_i}m_i^0/\gamma_{p_i}$, $i=1,2$.
From these equations we see that the ratio $R\equiv p_2/p_1$ is not any more fixed but takes the value:
\begin{equation}
R\equiv \frac{p_2}{p_1} = \frac{p_2^0}{p^0_1}\frac{1}{1+\eta^n} 
\frac{1+\frac{\theta^{free}_1}{\lambda_1}}{1+\frac{\theta^{free}_2}{\lambda_2}}
\end{equation}


It would be interesting to obtain the same ratio as a function of $M_{tot}$ instead of $M_{free}$.

$M_{tot}$ can be easily obtained from $M_{free}$, $m_1$ and $m_2$
\begin{equation}
\begin{array}{rl}
M_{tot} =& M_{free}\left(1+\frac{m_1}{\lambda_1}+\frac{m_2}{\lambda_2}+\frac{\gamma_s}{\alpha}\right)+ \frac{k_s}{\alpha}\\
\alpha =& k_1^{off}+k_2^{off}+\gamma_{c_1}+\gamma_{c_2}\\
\end{array}
\end{equation}

\begin{figure}[h]
	\centering
	\includegraphics{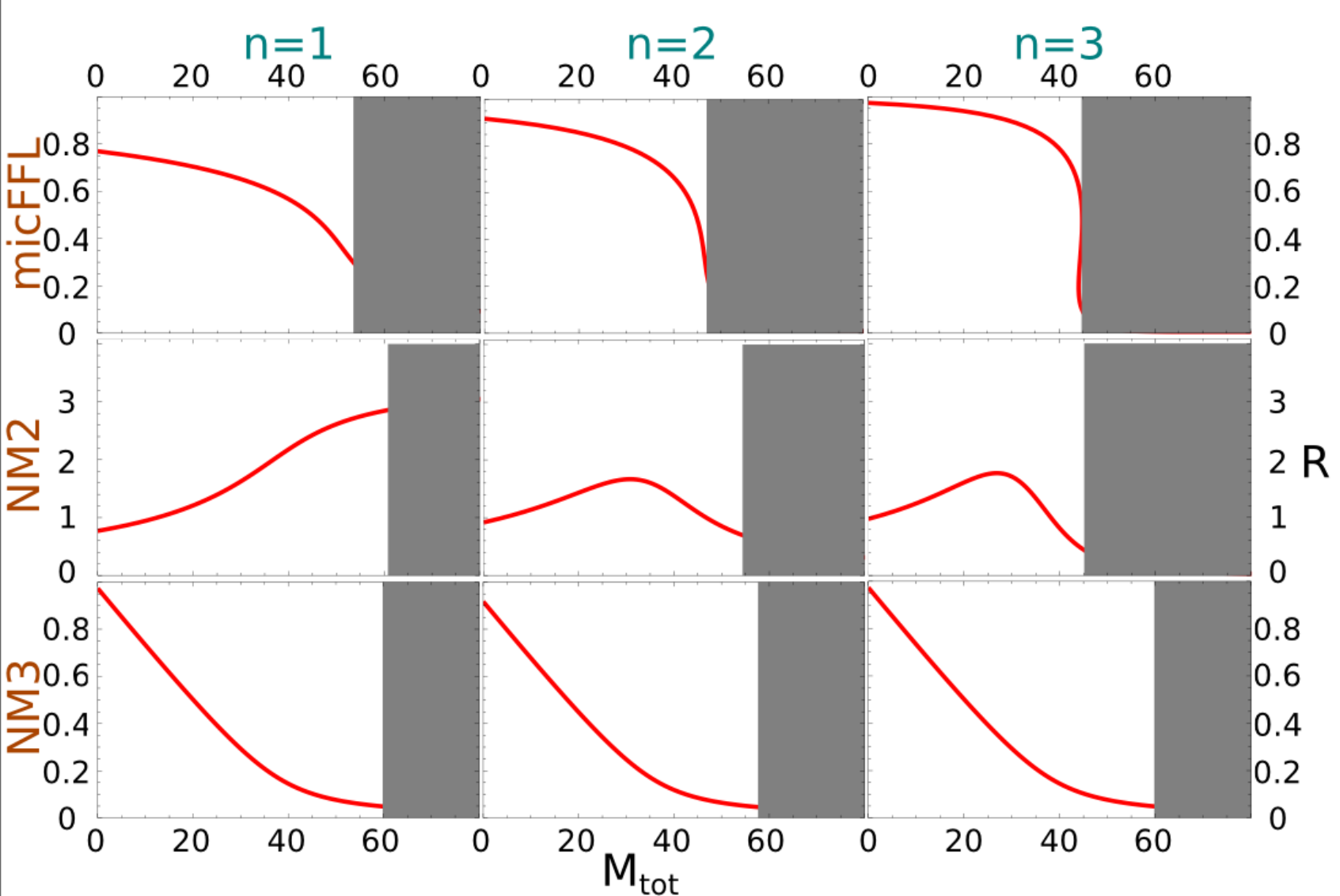}
	\caption{\label{fig:shapes}}
\end{figure}

The dependence on $m_1,m_2$ makes it difficult to write the ratio explicitly in terms of $M_{tot}$, but it can be easily obtain numerically.
We plot $R$ as a function of $M_{tot}$ in Figure \ref{fig:shapes}  in the limit in which $\theta_1^{free}=\theta_2^{free}\equiv\theta^{free}$ and $\lambda_1=\lambda_2=\lambda$ for $n=1,2$ and $3$ . We also plot for comparison the same ratio for the two null models NM2 and NM3. The shadowed portions of the plots denote the regions in which either $p_1/p_1^0$ or $p_2/p_2^0$ is less than 0.05, i.e the region in which the miRNA concentration is so high that one of the  proteins (or both) is almost absent and it would be meaningless to define a concentration ratio. Looking at the figures we see that, as the miRNA concentration increases, $R$ can be tuned from $p_2^0/p_1^0$ down to less than $20\%$ of the orginal value. The shape of the $M_{tot}$ dependence and the minimum value of  $R$ which can be reached strongly depend on the Hill coefficient. It is interesting to observe that also the two other motifs involving miRNA interactions that we studied (NM2 and NM3) allow to fine tune $R$ essentially to any desired value. These two models represent the limiting situations which one would obtain when $\lambda_1>>\lambda_2$ or $\lambda_1<<\lambda_2$.



\subsection*{Stochastic Analysis}

As in the previous sections we assume a titrative form for the miRNA-target interaction and an activatory Hill function to describe the TF dependent target transcription rate.
The molecular species we considered are transcripts for miRNAs ($s$), transcription factor ($m_1$) and target ($m_2$); proteins for transcription factor ($p_1$) and target ($p_2$), and the complexes that the miRNA can form when it binds to its target mRNAs ($c_1$ and $c_2$ respectively).
The chemical reactions involved in the circuit are schematically reported in Figure \ref{figFFL}C.
The corresponding master equation is:
\begin{equation}\label{eq:ME}
	\begin{array}{rl}
	\frac{\partial P(\{n_i\},t)}{\partial t}&=\left\{k_s(\mathcal{E}^{-1}_{1}-1)+\gamma_s(\mathcal{E}^1_1-1)n_1+k_{m_1}(\mathcal{E}^{-1}_2-1)+\right.\\
	&+\gamma_{m_1}(\mathcal{E}^1_2-1)n_2+k^{on}_1(\mathcal{E}^1_1\mathcal{E}^1_2\mathcal{E}^{-1}_3-1)n_1n_2+\\
	&+\alpha\gamma_{c_1}(\mathcal{E}^{-1}_1\mathcal{E}^1_3-1)n_3+k_{p_1}n_2(\mathcal{E}^{-1}_4-1)+\gamma_{p_1}(\mathcal{E}^1_4-1)n_4+\\
	&+k_{m_2}(\frac{TF^{ss}}{(TF^{ss})^n+h^n}-n\frac{\left(\frac{TF^{ss}}{h}\right)^n}{\left(1+\left(\frac{TF^{ss}}{h}\right)^n\right)^2}
	+\frac{n}{TF^{ss}}\frac{\left(\frac{TF^{ss}}{h}\right)^n}{\left(1+\left(\frac{TF^{ss}}{h}\right)^n\right)^2}n_4)(\mathcal{E}^{-1}_5-1)+\\
	&+\gamma_{m_2}(\mathcal{E}^1_5-1)n_5+k^{on}_2(\mathcal{E}^1_1\mathcal{E}^1_5\mathcal{E}^{-1}_6-1)n_1n_5+\alpha\gamma_{c_2}(\mathcal{E}^{-1}_1\mathcal{E}^1_6-1)n_6+\\
	&+k_{p_2}n_5(\mathcal{E}^{-1}_7-1)+\gamma_{p_2}(\mathcal{E}^1_7-1)n_7+(1-\alpha)\gamma_{c_1}(\mathcal{E}^1_3-1)n_3+\\
	&\left.+(1-\alpha)\gamma_{c_2}(\mathcal{E}^1_6-1)n_6\right\}P(\{n_i\},t)
	\end{array}
\end{equation}

with $\{s\to n_1,m_1\to n_2,c_1\to n_3,p_1\to n_4,m_2\to n_5,c_2\to n_6,p_2\to n_7\}$, and where 
$\mathcal{E}$ is the step-operator and can be written as $\mathcal{E}_j^k=\sum_{l=0}^\infty \frac{k^l}{l!}\frac{\partial^l}{\partial n_j^l}$. 

As in \cite{Osella11, Bosia12} in the above equation we linearized the Hill function around the steady state value $TF^{ss}$ (see SM for further details).

The analogous equations for the three null models are discussed in the SM. 

We are interested in evaluating:
\begin{description}
\item{i)} the linear correlation coefficient, which measures how much two variables are linearly dependent, 
$r_{xy}=\frac{<xy>-<x><y>}{\sigma_x\sigma_y}$ ; 
\item{ii)} the coefficient of variation, which is a measure of noise, 
$\eta_i=\frac{\sigma_i}{<x_i>}$ ; 
\end{description}
These quantities can be evaluated in general for any molecular species, but we shall in particular be interested in the linear correlation between $T$ and $TF$.


To estimate these quantities we need the first  two moments of the probability distribution $P(\{n_i\},t)$. 

Due to the complexity of the master equation this cannot be done analytically not even by linearizing the target transcription rate, thus we decided to approach the problem in the framework of the linear noise approximation \cite{vanKampen}. In this framework it is straightforward to obtain the covariance matrix of the system directly from its macroscopic description \cite{Elf03} and thus obtain approximate expressions for the first two moments of the probability distribution $P(\{n_i\},t)$. 

We also performed a set of Gillespie simulations on the model in order to quantify the error due to the linear noise approximations. Details on all these calculations can be found in the SM.

We made an effort to present all the results in terms of potentially measurable parameters, such as miRNA transcription rate $k_s$ and miRNA-target interaction strenght $ F=\frac{k^{on}_{c_i}}{\gamma_{s}\gamma_{m_i}}$ \cite{Levine07}, where $k^{on}_{c_i}$ is the rate of complex association, $\gamma_{s}$ and $\gamma_{m_i}$ are microRNA and messengers degradation rates respectively. In order to understand the peculiar properties of our circuit we compared it with the three null models NM3,NM4 and NM5.

Given the large number of free parameters, such a comparison is not straightforward. Our strategy was to maintain equal all the corresponding parameters in the four models and then compare all of them with the direct regulation (NM1), i.e. with the situation in which the miRNA is switched off.

\subsection*{miRNA-controlled feedforward loop increases TF-T statistical correlation}

\begin{figure}[h]
	\centering
	\includegraphics{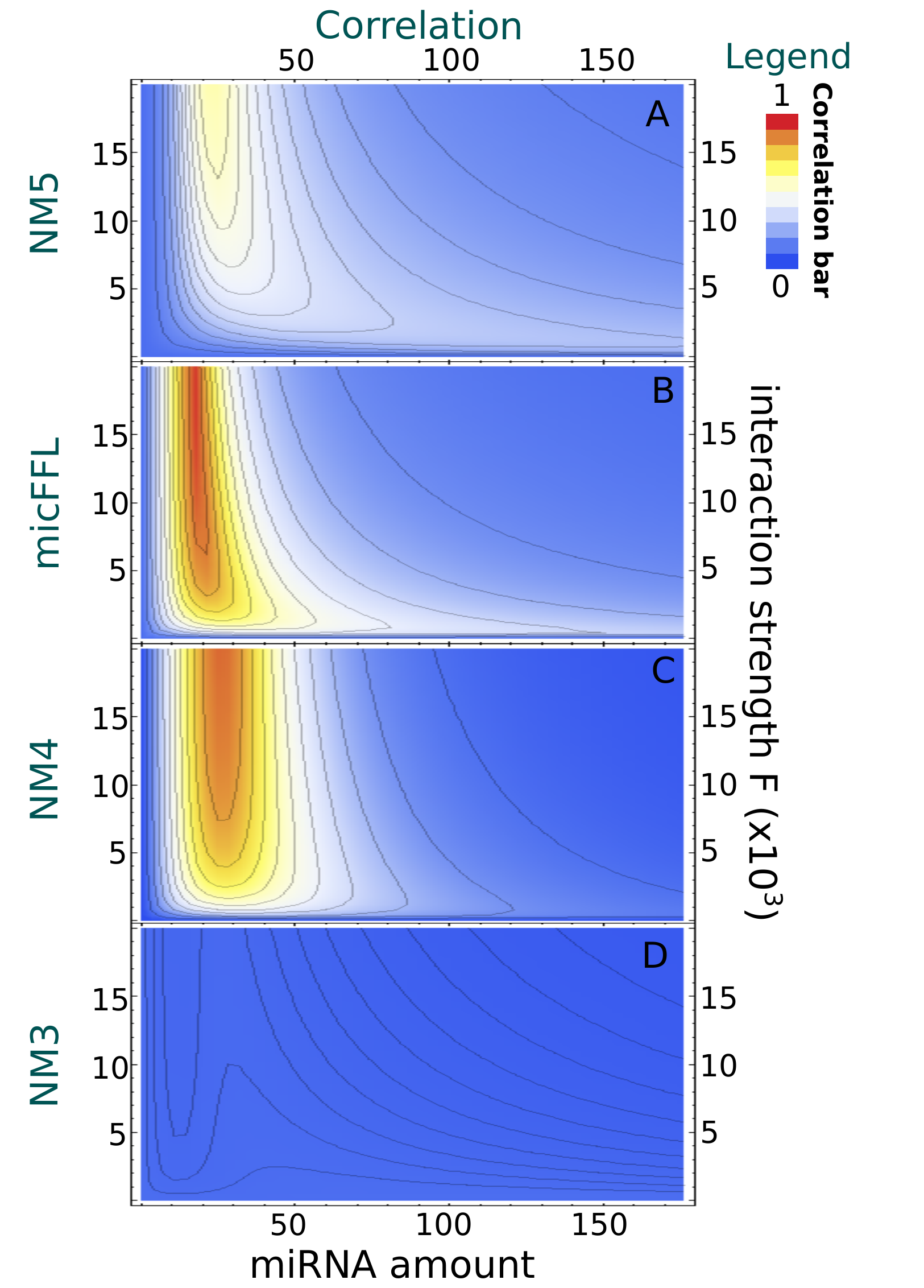}
	\caption{\label{fig:panel_corr_noise}}
\end{figure}

Our main result is the behaviour of the correlation coefficient between the Transcription Factor and its Target ($r_{TF,T}$). Our results are reported in Figure \ref{fig:panel_corr_noise}. We see that NM3 and NM5 show an almost negligible correlation while both in the micFFL and NM4 case there is a region of the parameter space in which TF and T are strongly correlated. In the NM4 case this behaviour was discussed in detail in \cite{BosiaCerna12}. It is a direct consequence of the titrative interaction between the miRNA and its targets which establishes an indirect interaction between transcripts in competition for binding the same miRNA. We think that this enhanced statistical correlation of targets is the ultimate reason for the generic enrichement observed in \cite{gerstein2012architecture, Sun12} for this type of motif when the targets are in physical interaction among them and thus are likely to require stable stoichiometric ratios. 

Looking at Figure \ref{fig:panel_corr_noise}B we see that the same correlation is present also in the micFFLs and turns out to be further enhanced by the transcriptional  link between TF and T. It is likely that this is the reason for the strong additional enrichment that we observe when reshuffling the miRNA-target links of our micFFLs.

We saw in the previous section that titrative interaction gives rise to threshold effects among the interacting molecules  and, as discussed in \cite{vanoudenaarden2011}, system hypersensitivity in proximity to the threshold. In the particular cases of the micFFL and of NM4 this effect involves three molecular species simulateneously: miRNAs  TFs and Ts and this gives rise to a very peculiar behaviour. Let us discuss the two cases separately:
\begin{itemize}
\item
In NM4 , when the amount of miRNA is similar to the amount of mTF and mT, a small fluctuation in even only one of their concentrations could be enough to move the system in the protein expressed or repressed phase. Thus, right in this condition of near-equimolarity of competing species the system is hypersensitive in changing of control parameters as miRNA or targets transcription rates \cite{ala2013}. The threshold is indeed determined by the model kinetic parameters and in the limit of strong interaction strength (high value of F) can be located in $k_s \sim k_{m_1} + k_{m_2}$ \cite{Elf03, Levine07, BosiaCerna12}.

\item
In miFFLC the situation is similar, but the direct link between TF and T increases the effective target transcription rate thus shifting the threshold 
toward a miRNA transcription rate higher than in the NM4. As a consequence, also the hypersensitivity region shifts its right-boundary. 

Moreover, fluctuations of the master regulator propagate to downstream genes. Thus, any fluctuation in the miRNA concentration
affects mTF and mT free molecules amount in the same direction (which is opposite with respect to miRNA) and the direct TF-T link further increases 
such fluctuations in the mT concentration (and consequently the TF-T correlation). This explains why in the 
micFFL the maximal level of correlation is  higher than in NM4.
The level of TF-T correlation might be seen as the result of the contribution of  two terms: the direct link TF-T  and  
the indirect miRNA-mediated TF-T link .
Figure \ref{fig:panel_corr_noise}A reproduces the situation in which two independent miRNA genes (with the same kinetic parameters of miFFLC) target TF and T independently
(NM5). 
The TF-T correlation profile results here from the bare fact that TF is an activator of T (direct link). 
In the NM4 case instead (Figure \ref{fig:panel_corr_noise}C) could be a proxy for the indirect effect alone.
Notice that the union of NM5 and NM4 correlation profiles is indeed very similar to the miFFLC one. 
For completeness we also analyzed the case in which the link miRNA-TF is lacking (NM3). Here again the correlation profile is due only to the direct TF-T connection. 
The heat-map does not show appreciable differences exploring the parameter space and the TF-T correlation values are almost everywhere comparable with that 
of a simple direct regulation.

\end{itemize}

\section*{Discussion}
\subsection*{MicFFLs role in the regulatory network}
The main outcome of the analyses discussed in the previous sections is that miRNA-controlled feed-forward loops are able to fine-tune the TF/target ratio to any desired value as a function of the miRNA concentration and that the peculiar topology of the circuit ensures a remarkable stability of this ratio against stochastic fluctuations. These two effects can be traced back to the particular form of miRNA/Target interaction which we assumed, the so called titrative interaction \cite{vanoudenaarden2011} which induces an indirect TF/Target interaction (via sponge effect) which compete with the standard transcriptional regulation. This additional interaction is controlled by the miRNA concentration which can thus fine tune the TF/target ratio. The interplay between direct and indirect interactions results in a stronger TF/Target correlation available for a broader range of miRNA concentration and miRNA/targets interaction strengths with respect to any other topology involving the three players, as confirmed by the comparison with the various null models that we studied.
It is likely that such a peculiar property of micFFLs could be very useful when TF and target must keep fixed concentration ratios, for instance if they must interact at a given stochiometric ratio. This is the case for instance of TF/target pairs involved in switch like functions like those which control the processes of tissue differentiation or cell proliferation or  the case of pairs of Transcription Factors which cooperate in regulating the same target. Indeed it was observed in \cite{gerstein2012architecture} that micFFLs involving proximal and distal regulators acting on the same gene are strongly enriched in the human regulatory network (see fig 1b). At the same time it is clear that in the generic situation such a linkage between TF and target should be avoided since the typical outcome of  transcriptional regulation is that a small change in the regulator should induce a much larger response in the regulated gene. This explains why this motif shows a strong negative enrichment when we reshuffle the transcriptional links of the network. On the contrary, the strong positive enrichment that we observed when reshuffling the post-transcriptional side of the network suggests that inducing a robust and stable fine tuning of the TF/target ratio could instead be one of the most important roles of miRNAs in the regulatory network. 
In order to elucidate this point we performed two further analyses: a functional enrichment analysis of the micFFLs targets and a comparison of the TF/target pairs with the PrePPI database of protein-protein interactions.

\subsection*{Functional Enrichment}
We performed a functional analysis of the target gene list corresponding to the FFLs obtained with the JASPAR TFs list and validated by all 4 miRNA-target databases and, separately, of the FFLs obtained with the ENCODE TF list. We used DAVID algorithm \cite{Huang2009,Huang2009b}, a comprehensive set of functional annotation tools, to understand biological meaning behind large lists of genes. We searched for enrichment based on Gene Ontology terms, Kegg metabolic pathways and human deseases. We found for a few categories an impressive enrichement (Bonferroni corrected p-values below $10^{-30}$). Remarkably enough the two lists of FFLs showed similar enrichment patterns and the most enriched categories turned out to be exactly the expected ones: Regulation of transcription, regulation of cell proliferation, positive regulation of cell differentiation, cell cycle and pathways in cancer. We report in Supplementary Table S1 (for the Jaspar list) and S2 (for the Encode list) the complete list of enriched categories with a False Discovery Ratio below $10^{-4}$.

\subsection*{MicFFLs with experimentally validated interactions}
In order to decrease the number of false positives in the list of putative micFFLs that we obtained with our bioinformatic analysis we selected among them those for which each one of the three regulatory interactions was experimentally validated in at least one experiment. This does not automatically means that all the three interactions are present in the same biological conditions and that the circuit is effectively active but it is certainly a strong indication in this direction. This list was obtained combining information collected from several databases (see details in the Material and Method section). We obtained in this way  a list of 499 micFFLs involving 365 distinct TF-target pairs  which are reported in the Supplementary Table S3,S4 and S5. We consider this list as our best candidates for a possible experimental validation of the micFFL's properties that we discussed in the previous sections.

\subsection*{Comparison with the PrePPI database}
We tested our conjecture that micFFLs could have a role in stabilizing the stochiometric ratio of proteins involved in physical interactions by comparing our list of micFFLs with experimentally validated interactions with the list of protein protein interactions collected in the PrePPi database \cite{PrePPI2012}.
Interactions in the database are validated through an algorithm based on 3d structure and functional analysis of the polypeptide chain. The algorithm was trained on the interactions of the major databases known till August 2010 and checked through the new interactions noted between august 2010 and august 2011. After training, Zhang's group predicted about 700 new interactions added to the PrePPI database.
We found that 30 out of the 499 pairs TF-target were present in the PrePPi database while the expected number was less than one. Assuming a binomial distribution we found a p-value of less than $10^{-50}$. While it is clear that we should consider this value with caution, since both our database and the PrePPi one contain experimentally validated data which are statistically biased, the gap between the number of expected interactions and those that we actually found is so large that it strongly supports our conjecture that micFFLs fine tune and stabilize the relative concentrations of interacting proteins.

\subsection*{Switch-on and switch-off response times}
\begin{figure}[h]
	\centering
	\includegraphics{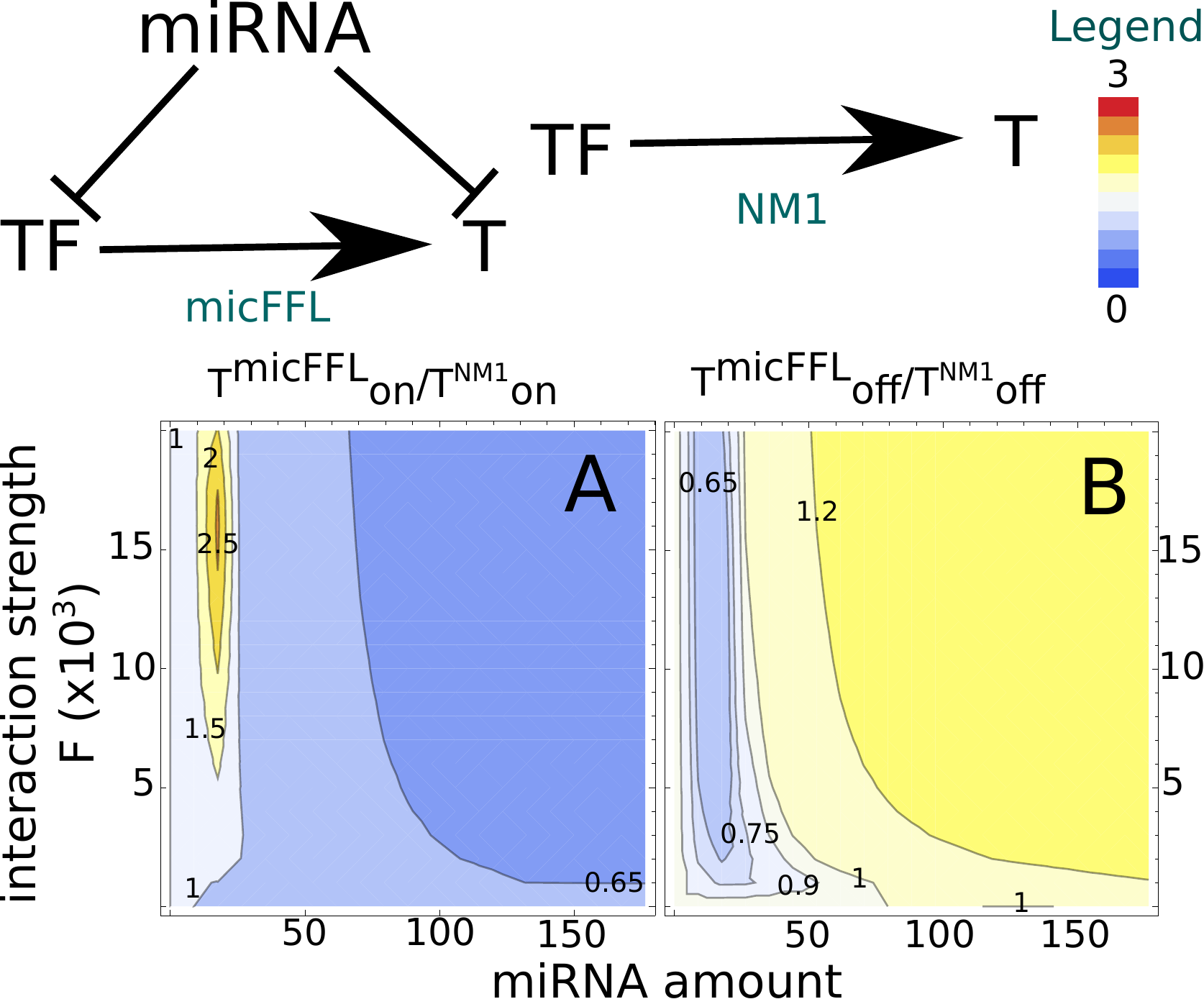}
	\caption{\label{RT}}
\end{figure}

In several cases the price one has to pay to be able to tightly control protein concentrations is a slowing down of response times. In order to better understand this issue we evaluated the switch-on and switch-off response times of the target in a micFFL and compared them with the analogous quantities in the case of a simple TF-target interaction, i.e. without the miRNA. We fixed the parameters of the micFFL so as to have the same steady state concentrations both for the TF and for the target. In this way, the only remaining free parameters are the concentration of the miRNA and its interaction strength, and we can study the change in the switch-on and switch-off response times as a function of these quantities. The results are reported in Figure \ref{RT}. As it is easy to see the response times are always of the same order of magnitude of the direct ones. In particular we see that as the miRNA concentration increases the switch-on time decreases and for physiological concentrations  the target in the micFFL reaches the steady state {\sl faster} than in absence of the miRNA. The efficiency of the miRNA plays only a minor role in this trend. The opposite is true for the switch-off time which shows a moderate increase while increasing miRNA concentartion and are instead strongly depressed for low miRNA concentrations.

\subsection*{A prototypical example: the micFFL  involving E2F1 and RB1 as targets and a set of miRNAs (miR-106a,miR-106b, miR-17, miR-20a and miR-23b) as master regulators}
\begin{figure}[h]
	\centering
	\includegraphics{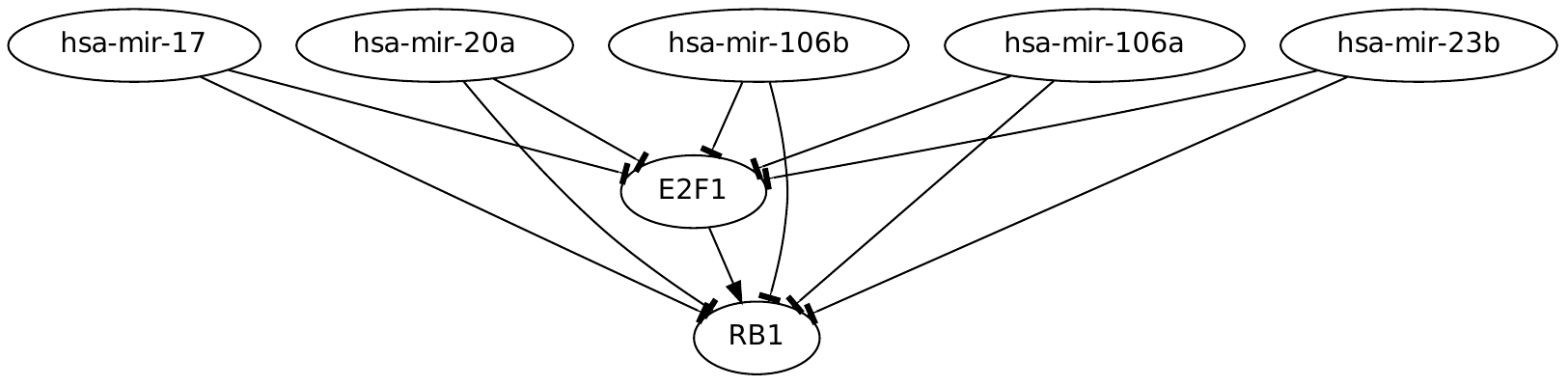}
	\caption{\label{fig:RB_network}}
\end{figure}

Within the list of candidates with experimentally validated interactions we selected, as an example, the micFFLs involving E2F1 and RB1 as targets and a set of miRNAs (miR-106a, miR-106b miR-17 miR-20a and miR-23b) as master regulators (see tab. S4). The network involving these genes is reported in Figure \ref{fig:RB_network}. The experimental support for these circuits is very strong see \cite{gerstein2012architecture} for the transcriptional regulation and \cite{Trompeter11} for those involving the miRNAs). E2F1 and RB1 are known to physically interact \cite{alberts,goodrich2006} and in fact they are included in the PrePPi database. The E2F1-RB1 system is a well known important switch in the cell cycle. E2F1 belongs to the family of E2F genes which  control the transition from G0/G1 to S phase in the cell (the quiescent phase and the first checkpoint phase respectively). In absence of mitogenic stimulation, E2F-dependent gene expression is inhibited by interaction between E2F and members of the retinoblastoma protein family RB (composed by RB1, RBL1 and RBL2), see \cite{goodrich2006}. When mitogens stimulate cells to divide, RB family members are phosphorilated then reducing their binding to E2F. The thus free-from-binding E2F proteins in turn activate expression of their target genes and trigger cell cycle. In G0 phase almost all cells have E2F1 and RB1 proteins bound in complexes \cite{alberts,goodrich2006}. In this state RB stops E2F functions and consequently the cell cycle. It is clear that the fine tuning and the stability of the relative concentration of the two genes is of crucial importance for the correct functioning of this checkpoint. Our analysis suggests that the stability of the checkpoint against stochastic fluctuations is guaranteed by the five miRNAs listed above and by the peculiar topology of the micFFLs which they form with their targets. The fact that the E2F1-RB1 pair is targeted simultaneously by five miRNAs is likely to reinforce the stabilization function. In our databases there are several other instances of TF-target pairs targeted by more than one miRNA. These are most probably the best candidates for further theoretical and experimental studies.

\section*{Materials and methods}

\subsection*{Construction of the post-transcriptional side of the regulatory network}

As potential targets of miRNAs we selected only transcripts corresponding to  protein-coding genes completely annotated in Ensembl 68 \cite{Flicek2012}, for a total of 76722 known transcripts. To define miRNA targets we used four freely available databases, chosen so as to have the widest possible spectrum of different prediction strategies. Three of them: 
doRiNA \cite{Anders2012}, microRNA.org \cite{Betel2008} and TargetScan \cite{Lewis2003} use algorithms based on sequence search similarity, possibly considering target site evolutionary conservation, the last one: PITA \cite{Kertesz2007} uses an algorithm based on thermodynamic stability of the RNA-RNA duplex, considering free energy minimization.
Integrating  the four databases we found a total of 4638441 interactions, involving 1581 miRNAs. For each miRNA-target link 
we annotated how many databases confirm the interaction. Then, out of these interactions, we selected those involving only 
Transcription Factors as targets. We based our analysis on two different TFs databases: JASPAR \cite{Stormo2000,Wasserman2004} and ENCODE \cite{gerstein2012architecture}. We found 34614 miRNA-TF interactions interactions for the JASPAR list, comprising 127 TFs, and 39498 for ENCODE  list, involving 121 TFs.

\subsection*{Construction of the transcriptional side of the regulatory network}

TF-target interactions were obtained with two different strategies depending on the TF database. For the JASPAR \cite{Stormo2000,Wasserman2004} TFs list we made use of the Position Frequency Matrix (PFM) information contained in the database \cite{Wasserman2004} and constructed a standard search algorithm for transcription factor binding sites (TFBS) within the target promoter region. Following the same procedure we adopted in our previous works on the subject (see for instance \cite{Re09,Friard2010}) we choose 1kb long  promoter regions, from 900 bases before the transcription start site (TSS) to 100 bases after the TSS. We used the scoring function proposed in \cite{Wasserman2004}, setting as threshold 0.70 of the max score. We found in this way a total of 948125 interactions. For the ENCODE TFs list  we used the ChIP-seq data obtained within the framework of the ENCODE project \cite{gerstein2012architecture}. These data were obtained for the 121  TFs over 5 main cell lines. We combined together the results of the different cell lines obtaining a total of 45328 TF-target interactions. 

\subsection*{Identification of micFFLs}

We constructed the list of putative micFFL by simply combining the interaction links obtained as discussed above. 
For the JASPAR list we obtained a total of 75933600 circuits, while using the data of the ENCODE project we obtained 2426300 micFFLs. In order to reduce the number of false positives we then selected only the micFFL with  both miRNA regulatory links confirmed by all the four databases. We obtained in this way 129100 micFFLs in the Jaspar case and 3782 in the Encode case. The list of these micFFL is available upon request.

\subsection*{Identification of micFFLs with experimentally validated regulatory interactions}

The list of micFFLs with experimentally validated regulatory interactions was obtained combining information collected from several databases.
\begin{itemize}
\item
For the miRNA $\rightarrow$ Target and the miRNA $\rightarrow$ TF interactions we used the last versions of the following miRNA databases: miRTarBase V 3.5 (updated November, 2012), miRecords V.3 (updated on November, 2010) and miR2Disease (updated on Jun, 2010). We obtained in this way a list of experimentally validated miRNA target gene interactions, containing 462 miRNAs, 2,280 target genes and a total of 4,277 independent interactions in human.
\item
For TF $\rightarrow$ Target interactions we used data from Encode \cite{gerstein2012architecture} (which contains a total of 44,842 regulatory interactions involving 122 TFs and 10,104  target genes) and the last version of Tfact(v.2) that contains genes responsive to transcription factors, according to experimental evidence reported in literature. It reports two datasets: (i) a sign sensitive catalogue that indicates the type (up or down) of TF regulation exerted on its targets; 
(ii) a sign less catalogue that includes all regulatory interactions contained in sign sensitive and further interactions 
without the specific type of regulation. Focusing on human only, the database contains a total 4,299 regulatory interactions involving 276 TFs and 1,937 target genes. The total number of non-redundant  TF-target regulatory interactions obtained combining the two datasets is 48,850 with 335 TFs and 10,828 target genes.
\end{itemize}
Combining the two datasets we obtained a total of 499 micFFLs. Out of them 95 involved a target which was itself a TF and for 7 of them (reported in supplementary Supplementary Table S5) the transcriptional regulation was bidirectional, in the remaining 88 (reported in Supplementary Table S4)instead only one of the TFs regulates the other one and there is no reciprocal interaction. Finally, in the remaining 404 micFFLs (reported in Supplementary Table S3) the target was not a TF. 

\subsection*{Simulations and Analytic calculations}

Analytical results have been obtained with Mathematica 8.0. 
Simulations present in SM have been obtained implementing Gillespie's direct algorithm \cite{gillespie1976}.

\section*{Acknowledgements}
We thank M. Cosentino Lagomarsino, M. Osella and D. Remondini for useful comments and suggestions.

\bibliographystyle{plos2009}
\bibliography{plos}

\begin{thebibliography}{10}
\providecommand{\url}[1]{\texttt{#1}}
\providecommand{\urlprefix}{URL }
\expandafter\ifx\csname urlstyle\endcsname\relax
  \providecommand{\doi}[1]{doi:\discretionary{}{}{}#1}\else
  \providecommand{\doi}{doi:\discretionary{}{}{}\begingroup
  \urlstyle{rm}\Url}\fi
\providecommand{\bibAnnoteFile}[1]{%
  \IfFileExists{#1}{\begin{quotation}\noindent\textsc{Key:} #1\\
  \textsc{Annotation:}\ \input{#1}\end{quotation}}{}}
\providecommand{\bibAnnote}[2]{%
  \begin{quotation}\noindent\textsc{Key:} #1\\
  \textsc{Annotation:}\ #2\end{quotation}}
\providecommand{\eprint}[2][]{\url{#2}}

\bibitem{Martinez09}
Martinez NJ, Walhout AJM (2009) {The interplay between transcripton factors and
  microRNAs in genome-scale regulatory networks}.
\newblock Bio Essays 31: 435-445.
\bibAnnoteFile{Martinez09}

\bibitem{Alon07}
Alon U (2007) Network motifs: theory and experimental approaches.
\newblock Nature Rev Genet 8: 450-461.
\bibAnnoteFile{Alon07}

\bibitem{Re09}
Re A, Cor\'a D, Taverna D, Caselle M (2009) {Genome-wide survey of
  microRNA-transcription factor feed-forward regulatory circuits in human}.
\newblock {Molecular BioSystems} 5: 854-867.
\bibAnnoteFile{Re09}

\bibitem{Shalgi07}
Shalgi R, Lieber D, Oren M, Pilpel Y (2007) {Global and local architecture of
  the mammalian microRNA-transcription factor regulatory network}.
\newblock {PLoS Computational Biology} 3(7): e131.
\bibAnnoteFile{Shalgi07}

\bibitem{Tsang07}
Tsang J, Zhu J, van Oudenaarden A (2007) {MicroRNA-mediated feedback and
  feedforward loops are recurrent network motifs in mammals}.
\newblock {Mol Cell} 26: 753-767.
\bibAnnoteFile{Tsang07}

\bibitem{Yu08}
Yu X, lin J, Zack DJ, Mendell JT, Qian J (2008) {Analysis of regulatory network
  topology reveals functionally distinct classes of microRNAs}.
\newblock {Nucleic Acids Research} 36: 6494 -- 6503.
\bibAnnoteFile{Yu08}

\bibitem{Hornstein06}
Hornstein E, Shomron N (2006) {Canalization of development by microRNAs}.
\newblock {Nature Genetics} 38: S20-S24.
\bibAnnoteFile{Hornstein06}

\bibitem{Osella11}
Osella M, Bosia C, Cor{\'a} D, Caselle M (2011) {The role of incoherent
  microRNA-mediated feedforward loops in noise buffering}.
\newblock {PLoS Computational Biology} 7: e1001101.
\bibAnnoteFile{Osella11}

\bibitem{Bosia12}
Bosia C, Osella M, El~Baroudi M, Cor\'a D, Caselle M (2012) {Gene
  autoregulation via intronic microRNAs and its functions}.
\newblock BMC Systems Biology 6: 131.
\bibAnnoteFile{Bosia12}

\bibitem{gerstein2012architecture}
Gerstein M, Kundaje A, Hariharan M, Landt S, Yan K, et~al. (2012) {Architecture
  of the human regulatory network derived from ENCODE data}.
\newblock Nature 489: 91--100.
\bibAnnoteFile{gerstein2012architecture}

\bibitem{Sun12}
Sun J, Gong X, Purow B, Zhao Z (2012) {Uncovering MicroRNA and Transcription
  Factor Mediated Regulatory Networks in Glioblastoma}.
\newblock {PLoS Comput Biol} 8(7): e1002488.
\bibAnnoteFile{Sun12}

\bibitem{arvey2010target}
Arvey A, Larsson E, Sander C, Leslie CS, Marks DS (2010) {Target mRNA abundance
  dilutes microRNA and siRNA activity}.
\newblock {Molecular Systems Biology} 6.
\bibAnnoteFile{arvey2010target}

\bibitem{Salmena2011}
Salmena L, Poliseno L, Tay Y, Kats L, Pandolfi PP ({2011}) {A ceRNA Hypothesis:
  The Rosetta Stone of a Hidden RNA Language?}
\newblock {Cell} {146}: {353-358}.
\bibAnnoteFile{Salmena2011}

\bibitem{Sumazin2011}
Sumazin P, Yang X, Chiu HS, Chung WJ, Iyer A, et~al. ({2011}) {An Extensive
  MicroRNA-Mediated Network of RNA-RNA Interactions Regulates Established
  Oncogenic Pathways in Glioblastoma}.
\newblock {Cell} {147}: {370-381}.
\bibAnnoteFile{Sumazin2011}

\bibitem{vanoudenaarden2011}
Mukherji S, Ebert M, Zheng G, Tsang J, Sharp P, et~al. (2011) {MicroRNAs can
  generate thresholds in target gene expression}.
\newblock {Nature Genetics} 43: 854--859.
\bibAnnoteFile{vanoudenaarden2011}

\bibitem{Elf03}
Elf J, Paulsson J, Berg O, Ehrenberg M (2003) Near-critical phenomena in
  intracellular metabolite pools.
\newblock {Biophysical Journal} 84: 154--170.
\bibAnnoteFile{Elf03}

\bibitem{Levine07}
Levine E, Zhang Z, Kuhlman T, Hwa T (2007) {Quantitative characteristics of
  gene regulation by small RNA}.
\newblock {PLoS Biology} 5: e229.
\bibAnnoteFile{Levine07}

\bibitem{Mitarai07}
Mitarai N, Andersson AMC, Krishna S, Semsey S, Sneppen K (2007) {Efficient
  degradation and expression prioritization with small RNAs}.
\newblock {Physical Biology} 4: 164.
\bibAnnoteFile{Mitarai07}

\bibitem{BosiaCerna12}
Bosia C, Pagnani A, Zecchina R (2013) {Modelling competing endogenous RNA
  networks}.
\newblock {PloS ONE} 8 (6).
\bibAnnoteFile{BosiaCerna12}

\bibitem{Figliuzzi12}
Figliuzzi M, De~Martino A, Marinari E (2013) {MicroRNAs as a selective,
  post-transcriptional channel of communication between ceRNAs: a steady-state
  theory}.
\newblock {Biophysical Journal} 104: 1203-13.
\bibAnnoteFile{Figliuzzi12}

\bibitem{Anders2012}
Anders G, Mackowiak SD, Jens M, Maaskola J, Kuntzagk A, et~al. ({2012})
  {doRiNA: a database of RNA interactions in post-transcriptional regulation}.
\newblock {Nucleic Acids Research} {40}: {D180-D186}.
\bibAnnoteFile{Anders2012}

\bibitem{Betel2008}
Betel D, Wilson M, Gabow A, Marks DS, Sander C ({2008}) {The microRNA.org
  resource: targets and expression}.
\newblock {Nucleic Acids Research} {36}: {D149-D153}.
\bibAnnoteFile{Betel2008}

\bibitem{Lewis2003}
Lewis B, Shih I, Jones-Rhoades M, Bartel D, Burge C ({2003}) {Prediction of
  mammalian microRNA targets}.
\newblock {Cell} {115}: {787-798}.
\bibAnnoteFile{Lewis2003}

\bibitem{Kertesz2007}
Kertesz M, Iovino N, Unnerstall U, Gaul U, Segal E ({2007}) {The role of site
  accessibility in microRNA target recognition}.
\newblock {Nature Genetics} {39}: {1278-1284}.
\bibAnnoteFile{Kertesz2007}

\bibitem{Flicek2012}
Flicek P, Amode MR, Barrell D, Beal K, Brent S, et~al. ({2012}) {Ensembl 2012}.
\newblock {Nucleic Acids Research} {40}: {D84-D90}.
\bibAnnoteFile{Flicek2012}

\bibitem{Stormo2000}
Stormo G ({2000}) {DNA binding sites: representation and discovery}.
\newblock {Bioinformatics} {16}: {16-23}.
\bibAnnoteFile{Stormo2000}

\bibitem{Wasserman2004}
Wasserman W, Sandelin A ({2004}) {Applied bioinformatics for the identification
  of regulatory elements}.
\newblock {Nature Review Genetics} {5}: {276-287}.
\bibAnnoteFile{Wasserman2004}

\bibitem{vanKampen}
Van~Kampen NG (2007) Stochastic processes in physics and chemistry.
\newblock Elsevier Science \& Technology Books, 3 edition.
\bibAnnoteFile{vanKampen}

\bibitem{ala2013}
Ala U, Karreth FA, Bosia C, Pagnani A, Taulli R, et~al. (2013) {Integrated
  transcriptional and competitive endogenous RNA networks are cross-regulated
  in permissive molecular environments}.
\newblock PNAS 110 (18): 7154-7159.
\bibAnnoteFile{ala2013}

\bibitem{Huang2009}
Huang DW, Sherman BT, Lempicki RA ({2009}) {Systematic and integrative analysis
  of large gene lists using DAVID bioinformatics resources}.
\newblock {Nature Protocols} {4}: {44-57}.
\bibAnnoteFile{Huang2009}

\bibitem{Huang2009b}
Huang DW, Sherman BT, Lempicki RA ({2009}) {Bioinformatics enrichment tools:
  paths toward the comprehensive functional analysis of large gene lists}.
\newblock {Nucleic Acids Research} {37}: {1-13}.
\bibAnnoteFile{Huang2009b}

\bibitem{PrePPI2012}
Zhang QC, Petrey D, Deng L, Qiang L, Shi Y, et~al. ({2012}) {Structure-based
  prediction of protein-protein interactions on a genome-wide scale}.
\newblock {Nature} {490}: {556+}.
\bibAnnoteFile{PrePPI2012}

\bibitem{Trompeter11}
Trompeter HI, Abbad H, Iwaniuk KM, Hafner M, Renwick N, et~al. (2011)
  {MicroRNAs MiR-17, MiR-20a, and MiR-106b Act in Concert to Modulate E2F
  Activity on Cell Cycle Arrest during Neuronal Lineage Differentiation of
  USSC}.
\newblock {PLoS ONE} 6(1): e16138.
\bibAnnoteFile{Trompeter11}

\bibitem{alberts}
Alberts (2008) {Molecular Biology of the Cell}.
\newblock {Garland Science}, 5 edition.
\bibAnnoteFile{alberts}

\bibitem{goodrich2006}
Goodrich D (2006) The retinoblastoma tumor-suppressor gene, the exception that
  proves the rule.
\newblock Oncogene 25: 5233--5243.
\bibAnnoteFile{goodrich2006}

\bibitem{Friard2010}
Friard O, Re A, Taverna D, De~Bortoli M, Cora D ({2010}) {CircuitsDB: a
  database of mixed microRNA/transcription factor feed-forward regulatory
  circuits in human and mouse}.
\newblock {BMC Bioinformatics} {11}.
\bibAnnoteFile{Friard2010}

\bibitem{gillespie1976}
Gillespie D (1976) A general method for numerically simulating the stochastic
  time evolution of coupled chemical reactions.
\newblock {Journal of Computational Physics} 22: 403--434.
\bibAnnoteFile{gillespie1976}

\end{thebibliography}

\newpage

\section*{Figure legends}

\paragraph{Figure 1.} \textbf{A}. Schematic description of the circuits discussed in the paper: NM1: direct regulation; NM2: open motif in which the 
microRNA regulates only the transcription factor; 
NM3: open motif in which the microRNA regulates only the target; NM4: Open motif in which the microRNA regualtes both the TF and the target but the 
TF-target link is missing; NM5, open motif in which two different microRNAs regulate separately the TF and the target. In the box we show the activactory micFFL whose
deterministic and stochastic behaviour we studied in the paper.
\textbf{B}. Schematic view of the general miRNA controlled Feed Forward Loops (combining both activactory and repressive TF-target interactions) mined 
in the bioinformatic analysis discussed in the paper. \textbf{C}. Schematic description of the chemical reactions which must be taken into account to describe  
the miRNA-mediated feedforward loop with a miRNA-target titrative interaction.

\paragraph{Figure 2.} Randomization of miRNA-target links. Distribution of the number of FFLs for 1000 simulations obtained with JASPAR TFs 
list and confirmed by at least 4 miRNA databases (Z = 49,4).

\paragraph{Figure 3.} Randomization of miRNA-target links. Distribution of the number of FFLs for 1000 simulations obtained with ENCODE TFs 
list and confirmed by at least 4 miRNA databases (Z = 23,3).

\paragraph{Figure 4.} Randomization of TF-target links. Distribution of the number of FFLs for 1000 simulations obtained with JASPAR TFs 
list and confirmed by at least 4 miRNA databases (Z = -20,8).

\paragraph{Figure 5.} Randomization of TF-target links. Distribution of the number of FFLs for 1000 simulations obtained with ENCODE TFs 
list and confirmed by at least 4 miRNA databases (Z = -18,1).

\paragraph{Figure 6.} Steady state analysis with the logic approximation of the micFFL. Plots \textbf{A} and \textbf{B} show the mRNA concentrations, respectively, of
transcription factor ($m_1$) and target ($m_2$) as a function of the  microRNA concentration ($\theta$) in the limit $\lambda\to0$. 
$H_s$ represents the activation threshold of the Heaviside function.

\paragraph{Figure 7.} The ratio of the target and TF concentrations as a function of $M_{tot}$ for the micFFL and the NM2 and NM3 null models for three values $n=1,2$
and $3$ of  the Hill exponent.

\paragraph{Figure 8.} Heat map of the correlation $r_{TF,T}$ for the micFFL and NM3,NM4 and NM5 Null Models. In each plot the values of $r_{TF,T}$ 
is mapped as a function of the miRNA concentration and of the interaction strength $F$. While for NM3 and NM5  the fluctuation of TF and T are almost
uncorrelated, both NM4 and the micFFL show a well defined region of large correlation. This correlation occurs for rather low miRNA concentrations and for 
almost any value
of the miRNA-mRNA interaction strength.

\paragraph{Figure 9.} Comparison of switch-on (\textbf{A}) and switch-off (\textbf{B}) response times between micFFL and direct regulation (NM1).

\paragraph{Figure 10.} The network of micFFLs involving E2F1 as transcription factor and RB1 as target. 

\newpage

\section*{Figures}

\end{document}